\documentclass[article]{aastex}
\begin{document}

   \title{A High Density Thin layer confining the H$\,$II region
          M$\,$42. HHT measurements\footnotemark[0]}

   \author{A. Rodr\'{\i}guez$-$Franco\altaffilmark{1}$\,$\altaffilmark{2}}
   \affil{Departamento de Matem\'atica Aplicada (Biomatem\'atica), 
          Secci\'on departamental de Optica, Escuela Universitaria de Optica.
          Universidad Complutense de Madrid. Av. Arcos de Jal\'on s/n. 
          E-28037 Madrid. Spain}
   
   \author{T.L.Wilson \altaffilmark{3}}
   \affil{MPIFR, Postfach 2024, D-53010 Bonn, Germany}
   
   \author{J. Mart\'{\i}n$-$Pintado, and A. Fuente}
   \affil{Observatorio Astron\'omico Nacional (IGN), Campus 
                  Universitario, 
                  Apdo. 1143, E-28800, Alcal\'a de Henares, Spain}
		  
	\altaffiltext{1}{Observatorio Astron\'omico Nacional (IGN), Campus 
                  Universitario, 
                  Apdo. 1143, E-28800, Alcal\'a de Henares, Spain.}
\altaffiltext{2}{Present address: LMSA Project Office
National Astronomical Observatory of Japan
2-21-1 Osawa, Mitaka, Tokyo 181-8588 Japan.}
\altaffiltext{3}{Present address: Sub-mm Telescope Observatory, Steward Observatory, The 
                  University of 
                  Arizona, Tucson, AZ, 85721.}

 \footnotetext[0]{This work is based on 
measurements made with
the Heinrich Hertz Telescope, which is 
operated by the Submillimeter Telescope
Observatory on behalf of Steward Observatory 
and the Max-Planck-Institut f\"ur Radioastronomie. 
}

   \begin{abstract}
%______________________________________ Do not leave a blank line here!
%
We present HHT observations in the 
$N=3\rightarrow2$ rotational transition of the CN radical toward selected 
positions of
the Trapezium region and of the molecular Ridge in the Orion molecular cloud.
Two of the positions in the Ridge were also observed in the $N=2\rightarrow1$ 
line of CN and
$^{13}$CN.
The $N=3\rightarrow2$ CN lines have been combined with observations of the 
$N=2\rightarrow1$ and
$N=1\rightarrow0$ transitions of CN, and of the $N=2\rightarrow1$ of 
$^{13}$CN to estimate the physical conditions and CN abundances in the
molecular gas.
We analyze in detail the excitation of the CN lines and find that the 
hyperfine ratios
of the $N=3\rightarrow2$ line 
are always close to the Local Thermodynamic Equilibrium (LTE) optically thin 
values
even
in the case of optically thick emission. This is due to different excitation
temperatures for the different hyperfine lines.
From the line intensity ratios between the different CN transitions 
we derive H$_2$ densities of $\sim10^5\,$cm$^{-3}$ for the molecular Ridge
and of $\sim3\times10^6\,$cm$^{-3}$ for the Trapezium region. 
The CN column densities are one order of magnitude larger 
in the
Ridge than in the Trapezium region, but the CN to H$_2$ ratio is similar both 
in the
Trapezium and in the Ridge. 
The combination of the low CN column densities, high H$_2$ densities and 
relatively high CN abundances toward the Trapezium region requires that the CN 
emission arises from a thin
layer with a depth along the line of sight of only $\sim5\times10^{15}\,$cm.
This high density thin layer of molecular gas seems to be related with 
material that  confines the rear side of the
H$\,$II region Orion A. However the molecular layer is not moving as expected 
from the
expansion of the H$\,$II region, but it is 
``static'' with respect to the gas in the molecular cloud.
We discuss the implication of a high density ``static'' layer in the 
evolution of an H$\,$II region.  
\end{abstract}
%_____________________________________ Do not leave a blank line here!
 

%
%________________________________________________________________
\section{Introduction}

The dynamical interaction between newly formed massive
stars, their H$\,$II regions, and the surrounding neutral material has been a
topic of great interest.  Based on simple ideas, many have argued that H$\,$II
regions should expand quickly into the ambient molecular cloud, expanding
beyond the ultra-compact phase in 10$^2$--10$^3$~yr (e.g. Dreher \& Welch, 
1981,
Dreher et al. 1984).
It has been recognized that a lifetime paradox exists for
ultra-compact H$\,$II regions, in that there are far too many given the
current birth rate of massive stars (Wood \& Churchwell, 1989).

Numerous authors have suggested resolutions for the H$\,$II region lifetime
paradox (see review of Hollenbach et al. 1994): 
a) confinement by infalling material (e.g. Reid et al. 
1981),
b) champagne flows (e.g. Tenorio--Tagle, 1979), c) moving star bow
shock (e.g. Van Buren et al. 1990), d) disk photoevaporation model
(Hollenbach et al. 1994), and e) mass loaded winds (Dyson et al., 1995; Lizano 
et al.,
1996).  These models have been developed to various degrees
but all exhibit serious shortcomings when it comes to matching current
observational data. De Pree et al. (1995) have suggested
a simple resolution for the lifetime paradox.  
Revisiting the lifetime arguments,
but assuming confinement by hot (100$\,$K) and high density 
(10$^7\,$cm$^{-3}$)
ambient material, De Pree et al. (1995) find that 
the ultra-compact H$\,$II
region phase can persist for $\sim10^5\,$yr.

To study the interfaces between the H$\,$II
regions and the surrounding material, one must find the appropriate tracers.
The usual high density tracers such as CS are not useful because they are  
not specific to the interfaces of the H$\,$II regions: there are also
high column 
densities of this molecule in hot cores. 

From theoretical arguments and
observations it is known that the abundances of some molecules are enhanced 
by UV radiation from the H$\,$II region.
Based on the results of the Photon Dominated Regions (PDRs) in
NGC$\,$2023 and NGC$\,$7023 (Fuente et al. 1993, 1995) and on PDR chemical 
models 
(Sternberg and
Dalgarno 1995, Jansen et al. 1995),
we have selected CN as one of the best tracer 
of the material at the H$\,$II/H$_2$ boundary layer affected by the UV photons 
from the OB stars.

Rodr\'{\i}guez--Franco et al. (1998) have 
used the 30-m telescope to map the $N=2\rightarrow1$ and the $N=1\rightarrow0$ 
lines of CN towards M$\,$42 
and M$\,$43.
Modeling the CN emission, Rodr\'{\i}guez--Franco et al. (1998) have 
produced H$_2$ 
density maps in the Orion M$\,$42 and M$\,$43 regions. 
The CN emission reveals a number of high density 
($\sim 10^5\,$cm$^{-3}$) bars (north bar, optical bar,
molecular filaments, and molecular Ridge) that surround the 
Trapezium cluster. The CN bars represent the interfaces between the molecular 
cloud and
the major ionization fronts of M$\,$42, confining this H$\,$II region in 
basically all
directions except along the line of sight.
Surprisingly, the largest H$_2$ 
densities, 
$>10^6\,$cm$^{-3}$ are found towards the Trapezium region
where the emission from CO is relatively weak. 
This CN emission from the Trapezium
arises from the thin layer of molecular gas that is related with the 
material that interacts with the rear
ionization fronts of M$\,42$ confining the H$\,$II region in this direction.
The CN layer is likely to be the remnant of the high density layer 
($\sim10^7\,$cm$^{-3}$)
that also confined the H$\,$II region in the 
ultra- and hyper-compact phase.

The critical densities to excite the $N=1\rightarrow0$ and $N=2\rightarrow1$ 
lines
of CN are 
$\sim 10^6\,$cm$^{-3}$. Therefore from the analysis of these transitions one 
cannot
constrain the H$_2$ densities if they are larger than $10^6\,$cm$^{-3}$.
To estimate the H$_2$ densities in the Trapezium region one needs to measure a 
transition
with higher Einstein coefficients.
In this paper, we present observations of the CN 
$N=3\rightarrow2$
transition taken with an angular resolution similar to that of the 
$N=1\rightarrow0$
line 
and analyze the new results combined with 30-m data. The new data show that 
the H$_2$
densities in the Trapezium layer are $\sim 3\times10^6\,$cm$^{-3}$.

\section{Observations and results}
\label{sec:observations}

The observations of the $N=3\rightarrow2$ and the $N=2\rightarrow1$
lines of CN, at $340.2\,$GHz and $226.3\,$GHz respectively, 
and the $N=2\rightarrow1$ of $^{13}$CN at $217.4\,$GHz
were taken between 27 
April and 
3 May 1999 and April 2000 with the Heinrich Hertz Telescope (HHT) in Arizona 
(USA). At
the rest frequencies of the $N=3\rightarrow2$ and the $N=2\rightarrow1$, 
the full width to 
half power of the telescope was 26$''$ and 34$''$ respectively.
The pointing was checked by continuum cross scans of Venus. The RMS pointing 
error was
always less than $\sim$3$''$.
The 345$\,$GHz receiver is a facility instrument consisting of a dual channel 
superconducting mixer provided by 
the Max Plank Institut f\"ur Radioastronomie (MPIFR). 
The double sideband receiver noise temperatures were typically 
$\sim$150-200$\,$K and
$\sim$100$\,$K for 
the 345$\,$GHz and 230$\,$GHz receivers respectively.
At the elevation of Orion, the single sideband system noise temperature was 
typically 
$\sim$800 to 1400$\,$K for the 345$\,$GHz receiver and $\sim 390$ to $760\,$K 
for the
230$\,$GHz receiver.
The data were calibrated using the chopper wheel 
method. We have
used position switching, with the reference at ($-800''$, 0$''$) from the on-
source position.
We established the final temperature 
scale by comparisons of our Orion-KL results  with the data 
of Groesbeck et al. (1994) and Schilke et al. (1997) .

For some tunings of the 345$\,$GHz receiver, there is a frequency-dependent 
variable
ratio between the signal and image sidebands with a frequency of 
$\sim$500$\,$MHz.
To check for the presence of such variability, we shifted each spectrum five 
times by up to $300\,$km$\,$s$^{-1}$.
This also allowed us to measure all the hyperfine components
with the 
narrow band spectrometers, and to search for the presence of intense lines 
in the image spectrum.  

The spectra were analyzed using filter banks and 
Acoustic Optical Spectrometers (AOS's). The wide band AOS's with a bandwidth of 
983$\,$MHz
had a resolution 
of 480$\,$kHz and the narrow band AOS's with a bandwidth of 64$\,$MHz
has a resolution of 250$\,$kHz. The 
$F=9/2\rightarrow7/2$,
$F=7/2\rightarrow5/2$, and
$F=7/2\rightarrow3/2$ hyperfine components of the $N=3\rightarrow2$, 
$J=7/2\rightarrow5/2$ multiplet of CN are separated by 790$\,$MHz. This range 
of
frequencies
was in the analyzing band of the wide-band AOS's. For the $N=2\rightarrow1$ 
line of CN
the $F=5/2\rightarrow3/2$,
$F=3/2\rightarrow5/2$,
$F=3/2\rightarrow3/2$,
$F=3/2\rightarrow1/2$,
$F=1/2\rightarrow3/2$, and
$F=1/2\rightarrow1/2$
HF components
of the fine $J=3/2\rightarrow3/2$ transition was observed with the 64$\,$MHz 
narrow band
AOS.

The observations were made toward selected positions
in the Ridge and in the Trapezium region obtained from the CN 
$N=1\rightarrow0$ line
map of Rodr\'{\i}guez--Franco et al. (1998).
Fig. \ref{fig:mapa} shows superimposed on the CN $N=1\rightarrow0$ map from  
Rodr\'{\i}guez--Franco 
et al. (1998) the positions in the 
Trapezium region (T1\ldots T6) and in the molecular Ridge (R1\ldots R6) where 
we
took the spectra in the $N=3\rightarrow2$ line of CN. The $N=2\rightarrow1$ 
line of CN
and $^{13}$CN were measured toward the positions R4 and R5.

Fig. \ref{fig:espec} shows the spectra of
the $N=3\rightarrow2$ line toward all the observed positions, and 
Fig. \ref{fig:iso} the $N=2\rightarrow1$ spectra of CN and $^{13}$CN towards 
R4 and R5. Table 
\ref{tab:obs} gives the observational parameters derived
from Gaussian fits to the  $F=9/2\rightarrow7/2$,
$F=7/2\rightarrow5/2$, and
$F=7/2\rightarrow3/2$
hyperfine (hereafter HF) components of the 
$N=3\rightarrow2$, $J=7/2\rightarrow5/2$ multiplet.
Table \ref{tab:obs} gives also the observed parameters of the 
$N=2\rightarrow1$ line
of CN and $^{13}$CN the line towards R4 and R5 (see next section for a 
discussion).

\section{CN line opacities from the hyperfine ratios}

Usually one can use the intensity ratio between the fine and HF components of 
the
CN transitions
to estimate the optical depth of the lines (see e.g. Rodr\'{\i}guez--Franco et 
al. 1998).
This method is widely used for
molecules like NH$_3$ (Pauls et al. 1983) and gives reliable results when all 
the HF
components 
have the same
excitation temperature. In some cases, like for NH$_3$ and CN, it is well 
known
that the different HF components might have different excitation temperatures 
(Gaume et al. 1996).
We now  discuss  the problems associated with the
use the HF ratios to estimate the CN line opacities for the submillimeter 
lines of CN.

The $N=1\rightarrow0$ rotational
transition results in 9 HF components grouped in two main fine-structure
groups, the $N=2\rightarrow1$ and the $N=3\rightarrow2$ transitions
are split in 18 HF lines which are
grouped in three fine-structure groups, and the $^{13}$CN $N=2\rightarrow1$ 
line is
split into 71 HF
components.
In the wide band AOS we have observed the two most intense fine 
structure groups of
the $N=3\rightarrow2$ line simultaneously. We have also measured the 
$J=3/2\rightarrow3/2$
fine structure
group in the $N=2\rightarrow1$ line. 
We have derived the HF ratio for the observed CN lines by fitting a  ``comb'' 
of
Gaussian profiles centered at the frequencies and the relative intensities of 
the HF components 
for every CN line. Fig \ref{fig:iso} shows the fits of the HF components for 
the
CN and $^{13}$CN 
$N=2\rightarrow1$ line 
in the two positions of the Ridge. Similar good fits are also obtained for the 
CN
$N=3\rightarrow2$ HF components in Fig. \ref{fig:espec}. For both the Ridge 
and the Trapezium positions, we
derive  a ratio between the 
$J=7/2\rightarrow5/2$ and the $J=5/2\rightarrow3/2$ fine components of the 
$N=3\rightarrow2$ line
(RF32 hereafter) 
close to $\sim1.5$ (see Table \ref{tab:obs}), which would indicate 
optically 
thin emission. 
This is
surprising in view of the relatively strong CN $N=3\rightarrow2$ lines in the 
Ridge
and the detection of the $^{13}$CN $N=2\rightarrow1$
line towards two positions in the Ridge. 
These facts indicate that the CN lines in the Ridge cannot be optically thin 
in contrast
to the RF32 indicating optically thin emission. 
These
discrepancies are due to different excitation temperatures for different HF 
components.

%%%%%%%%%%%%%%%%%%tablas

In order to study  to what extent the assumption of equal excitation 
temperature for all
the HF components can be applied to the $N=3\rightarrow2$ line to 
derive the opacities, we have studied the
excitation of CN by combining
the CN and $^{13}$CN line intensities measured in this paper with those 
measured by
Rodr\'{\i}guez--Franco et al. (1998). For this, we have used the model based 
on the Large
Velocity Gradient (LVG)  approximation 
described by Rodr\'{\i}guez--Franco et al. (1998) and a constant kinetic 
temperature
of $80\,$K (Wilson et al. 2000). In this model only collisional 
excitation has been considered. The effects of the excitation of the first 
vibrationally excited levels by infrared radiation at 4.8 microns has been 
considered negligible for the typical physical conditions in photodissociation 
regions (Fuente et al. 1995). The data of Rodr\'{\i}guez--Franco et al. 
(1998) have been
smoothed to have the same angular resolution as the HHT CN $N=3\rightarrow2$ 
data.
Figs. \ref{fig:lvg}a and b show the results obtained from the LVG model 
calculations.
Fig. \ref{fig:lvg}a  shows the dependence of the H$_2$ densities (n$_{{\rm 
H}_2}$)
with the CN
column density (N(CN)) as a function of the intensity of main group of  
components of the 
$N=3\rightarrow2$ multiplet, $T_a^*$($3\rightarrow2$) (thick line) observed in 
this paper
and the ratio
between the intensities of the $N=3\rightarrow2$ and $N=2\rightarrow1$ lines, 
R31 (dotted
lines). In  Fig. \ref{fig:lvg}b  we plot the H$_2$ density versus the CN 
column
density as a function of the optical depth of the $N=3\rightarrow2$, 
$J=7/2 \rightarrow5/2$, $F=9/2 \rightarrow7/2$ line,
$\tau_{\rm m}(3\rightarrow2)$, and the ratio between the fine components RF32.
Fig. \ref{fig:lvg}b shows that model predicts for a rather wide range of H$_2$ 
densities and CN column densities, line 
intensity ratios around 1.5  (see Table 1). Even for the case of 
optically thick
emission with opacities between 1 and 5 the predicted value of RF32
will be of $\sim1.5$.
This is clearly the 
case for the two positions in the Ridge R4 and R5.

In order to fit the $^{13}$CN and CN lines simultaneously in the two Ridge 
positions
R4 and R5
where $^{13}$CN was observed, $^{13}$CN column  densities of 
$\sim10^{13}\,$cm$^{-2}$ are required. Assuming a standard CN/$^{13}$CN ratio 
of 89
(Wilson 1999), the CN column  density must be close to $\sim 
10^{15}\,$cm$^{-2}$. For this
column 
density, the HF component of CN $N=3\rightarrow2$ line presented in this paper 
is
optically thick with opacities $\leq5$
(see Fig. \ref{fig:lvg}b) and the expected HF ratio RF32 of $\sim1.5$. This 
indicates,
as previously
mentioned, that the HF ratio does not necessarily give a measure of the CN 
line optical
depths. 
This is because the excitation 
temperature is not the same for all the observed hyperfine components.
Since the excitation is
mainly collisional, the excitation temperature is expected to be higher in 
the hyperfine components with higher opacities. 
In fact, in the simple isothermal case considered in this paper,
the LVG model gives 
$T_{ex}\sim 23\,$K and $14\,$K for the 
$J=5/2\rightarrow3/2$ and $J=7/2\rightarrow5/2$ fine components of the 
$N=3\rightarrow2$
transition
respectively. Because of 
these 
differences in the excitation temperature, the ratio between the HF 
structure components
remains quite uniform and close to 1.5 for a large range of opacities. 

In summary, the HF line intensity ratio of the $N=3\rightarrow2$ CN line does 
not
necessary give a reliable estimate of the line opacity even if the ratios are 
close
to the expected
optically thin case. Observations of the $^{13}$CN line and/or a number of HF 
components
with a wide range of 
relative intensities are needed to estimate
the opacities of the  strongest HF lines of CN.

\section{Physical conditions}
As shown in Fig. \ref{fig:lvg}a one can use the line intensity ratio of the 
$N=3\rightarrow2$, and the $N=1\rightarrow0$ lines of CN to derive the H$_2$ 
density and the CN column density for the observed position in the Ridge and 
in the
Trapezium once the kinetic temperature is known.

In Table \ref{tab:derived} we present the results for the H$_2$ densities and 
CN column
densities for
the Trapezium and the Ridge for a kinetic temperature of $80\,$K. These 
results are basically independent of the assumed kinetic temperatures for 
kinetic temperatures larger than 25$\,$K (Rodr\'{\i}guez--Franco et al. 1998). 
For the positions
in the Ridge we cannot fit the intensity of the 
CN $N=3\rightarrow2$ line and the ratio R31 
for the optically thin solution. Therefore we have considered
the solution for the optically thick case found in the previous section for R4 
and R5.
For the
Trapezium positions, the CN $N=3\rightarrow2$ line is weaker than in the Ridge,
and the optically thin solution can explain both, the line intensities and the 
ratio R31.

The Trapezium and the Ridge regions present very different physical
conditions (see Table \ref{tab:derived}).
In the Trapezium region, the densities are between $6\times10^5$ and 
$7\times10^6\,$cm$^{-3}\,$cm$^{-3}$ 
and CN column densities around $4\times10^{13}\,$cm$^{-2}$. 
With these physical conditions
the opacity of the main hyperfine component of the CN $N=3\rightarrow2$
line is $\tau_m (3\rightarrow2)\leq0.5$ as expected from the hyperfine ratios 
of the
CN $N=3\rightarrow$2 and $N=1\rightarrow$0 lines. 
For the Ridge we obtain densities between $3\times10^{4}$ and 
$2\times10^{5}\,$cm$^{-3}$,
CN column 
densities between $3\times10^{14}$ and $2.5\times10^{15}\,$cm$^{-2}$, and 
opacities
between 1 and 5. 
The derived H$_2$ 
densities for the Ridge from the CN emission are in good agreement with
those derived from HC$_3$N 
(Rodr\'{\i}guez$-$Franco et al. 1992).

In summary, the $N=3\rightarrow2$ CN data
confirm the H$_2$ densities for the Trapezium region, typically of a few 
$10^6\,$cm$^{-3}$. These are
larger, on average, by a factor of $\sim10$ than those in the molecular Ridge
($0.3-2\times10^5\,$cm$^{-3}$).
However, the CN column  
densities in the Ridge ($\sim1\times10^{15}\,$cm$^{-2}$) are larger by
more than a factor of 10 than in the 
Trapezium ($\sim4\times10^{13}\,$cm$^{-2}$).

\subsection{CN abundances}
We have also derived the relative abundance of CN in both regions by combining 
the CN column
densities with the
H$_2$ column 
density estimated from the $^{13}$CO and C$^{18}$O data of White \& Sandell 
(1995) and
Wilson et al. (1986). The H$_2$ column  densities have been derived  
by using the LVG approximation to model the CO excitation 
and assuming a $^{13}$CO/C$^{18}$O ratio of 5.6. The results are given in Table
\ref{tab:derived}.
We find H$_2$ column densities of 
$0.5-2.4\times10^{22}\,$cm$^{-2}$ for the Trapezium region and of 
$4-6\times10^{23}\,$cm$^{-2}$ for the Ridge. 
>From the H$_2$ and CN column  densities, we estimate the CN/H$_2$ ratios shown
in column 5 of Table \ref{tab:derived}. The CN abundance ratios in the 
Trapezium region and molecular Ridge
are similar, although there is a systematic trend for the CN abundance to be 
somewhat
larger in the
Trapezium (average of $4\times10^{-9}$) than in the Ridge (average of 
$10^{-9}$).

\subsection{Depth along the line of sight of the CN emission}
Combining the H$_2$ densities derived for the multi-transitional analysis 
of CN with the
H$_2$ column densities derived from $^{13}$CO and C$^{18}$O,
we can also estimate the thickness of the molecular gas layer
along the line
of sight. In column 6 of 
Table \ref{tab:derived} we show the results. 
We derive a thickness for
the Ridge of typically $3\times10^{17}\,$cm. This is consistent with the 
emission arising from
spherical condensations with a size of $\sim5-10''$. The thickness of the
molecular layer in the Trapezium region is $10-100$ times smaller 
than the thickness of
the Ridge. The small thickness of the CN emission toward the Trapezium 
region makes it very 
unlikely that this emission arises from condensations with 
spherical geometry as for the Ridge, since this geometry would require
condensations with sizes of only $1''$. 
To explain the observed CN line intensities one would 
require a large number of these condensations filling the telescope beam. 
Since the CN emission is
extended we conclude that the CN emission in the Trapezium region arises 
from a thin,
$\sim5\times10^{15}\,$cm, layer of dense 
(few $10^{6}\,$cm$^{-3}$) molecular gas (hereafter the Trapezium 
molecular layer).

\section{Discussion}

Rodr\'{\i}guez--Franco et al. (1998) have discussed in detail the implications 
of
the CN emission in the confinement of the Orion A H$\,$II
region in the perpendicular direction to the line of sight. 
The geometry of the Trapezium molecular layer, 
and the  
chemistry, with large CN abundance as compared with other
molecules like HC$_3$N,
confirms that the CN emission is related with the material which 
confines the
rear side of Orion A.
The presence of a layer of material behind the Trapezium confining the 
Orion A 
H$\,$II region
is required to explain the  
kinematics of the ionized gas (Zuckerman, 1973). The confining material 
will show the typical multilayer (ionized, neutral atomic and molecular)  
structure expected when molecular material is exposed to UV radiation (see eg. 
Hollenbach \&Tielen 1997). Fig. \ref{fig:model} shows an
sketch of the distribution and kinematics of the ionized and
molecular gas towards the Trapezium. A model of the location of the 
different layers along the line of sight has been obtained by O'Dell et al. 
1993). The ionized gas is
blueshifted with respect to the neutral molecular gas,
indicating that the ionized gas is flowing towards the observer.  
To prevent the ionized gas from escaping from the H$\,$II region in the 
direction away from the observer, the rear side of the H$\,$II region must be 
confined
by a dense layer of gas and dust. This is confirmed by the radial 
velocities of O$\,$I and H$\,$I.
Both the atomic layer probed by O$\,$I and H$\,$I and the molecular 
layer probed by CN show velocities close to 9$\,$km$\,$s${-1}$, the
ambient cloud velocity, suggesting that this neutral gas forms the 
front face of the neutral confining cloud
(O'Dell et al. 1993; van der Wef \& Goss 1989).  The CN 
emission towards the Trapezium has similar radial velocities than that of the 
warm  CO (Howe et al. 1993; Wilson et al. 2000) heated by the UV 
radiation. As expected for a dense photodissociation region (Sternberg 
\& Dalgarno 1995), both molecular emission  arise from basically the same layer 
which located close to the principal ionization front (see eg. O'Dell et al. 
1993). At the opposite side of the ionization from the dense and warm 
molecular layer will be surrounding by an envelope of molecular and atomic 
material (see van der Wef \& Goss 1989).

The origin of the thin confining layer in the Trapezium and its
present morphology is unclear since the evolution of  H$\,$II regions is 
far from understood (see e.g. Churchwell 1990). 
One would think that the dense thin layer has been built up by 
material swept by the expansion of the H$\,$II region in the last $10^6\,$yr. 
In this case the molecular component of the confining material should be 
expanding. Expanding molecular gas with a  velocity of at least 
$5\,$km$\,$s$^{-1}$ has been observed from the CN emission in the  molecular
molecular bars (northern bar, optical filament) which are believed to belong 
to the material that confine the H$\,$II
region in the direction close to the plane of the sky (Rodr\'{\i}guez$-$Franco 
et al. 1998).
However,
the observe thin CN layer has a radial velocity of $\sim9\,$km$\,$s$^{-1}$,
typical of
the molecular gas unperturbed by the effects of the H$\,$II region.
These arguments point towards the possibility that either the CN Trapezium 
layer is not related to the material that
confines the H$\,$II region or to the fact that the evolution of M$\,$42 does 
not fit the
simple picture of an  H$\,$II region expanding at a velocity of 
$\sim15\,$km$\,$s$^{-1}$.
The first possibility is unlikely since the UV radiation seems to dominate the 
chemistry
of the thin molecular layer indicating small visual extinction 
($\leq4\,$mag) between the
ionizing front and the molecular layer (Rodr\'{\i}guez$-$Franco et al. 1998). 
Furthermore, the line widths of the CN are broader than those  found for the 
Ridge,
revealing a kinematic interaction 
between the H$\,$II
region and the Trapezium molecular layer. If the Trapezium molecular layer is 
indeed
interacting with
the H$\,$II region, the lack of velocity shifts between CO (from the bulk of 
the
Ridge) and CN from the Trapezium layer indicates that the H$\,$II region is 
not
expanding away from the observer.

The Trapezium molecular layer would therefore indicate that the presence of 
``static'' layers not only occurs at the early phases of the evolution of the
H$\,$II regions as shown by Wood \& Churchwell (1989), but also in more 
evolved
objects. Although several suggestions have been made to explain the lifetime, 
morphology
and kinematics of the ultra-compact
H$\,$II regions
(see e.g.Tenorio-Tagle 1979; Reid et al.
1980; Van Buren et al. 1990; Hollenbach et al. 1994;  Dyson et al. 1995; Lizano
et al. 1996),
none of these models adequately explain all the observational facts (Jaffe \& 
Mart\'{\i}n$-$Pintado, 1999). Most of the models developed to explain the 
lifetime
paradox such as the bow shocks, mass-loaded winds and confinement 
by infalling
material predict that the confining material is not ``static'' with respect to 
the
ambient molecular clouds. The disk evaporated
model cannot explain the morphology and kinematics of evolved H$\,$II regions 
such us
M$\,42$.

The champagne flow could explain both the kinematics of the 
ionized gas in M$\,$42 and the presence of the 
``static'' and the expanding molecular layers confining 
M$\,$42. In the scenario of the champagne model, the massive stars
are formed within a molecular cloud with a large density gradient.
The H$\,$II region confined by material with a density gradient will 
expand in some directions but depending on the density and temperature of of 
the
surrounding material, can be ``static'' in other directions.
As suggested by 
De Pree et al. (1995) very dense 
($10^7\,$cm$^{-3}$), and warm (100$\,$K) molecular material would
slow down the expansion of an H$\,$II region in the early phases
and  provide an explanation for the presence of ``static'' layers 
in more evolved H$\,$II regions. 
In the case of M$\,42$ the H$_2$ density gradient is along the line of 
sight with the larger density in the rear side of the H$\,$II region, just 
where the
Trapezium layer is found. In the Trapezium layer the 
H$_2$ density is larger than $10^6\,$cm$^{-3}$
and the kinetic temperatures are $\sim100\,$K (Wilson et al. 2000). The CN 
layer will be 
in pressure equilibrium with the 
ionized gas for the typical densities of a few $10^4\,$cm$^{-3}$ and
the electron temperature of $8000\,$K measured in M42 (Wilson et al. 1997).
The small thickness of the the CN layer, as compared to the Ridge,
can also be understood
in the framework of the champagne flow since the 
gas flowing towards the observer has been photo--evaporated 
from the confining layer as sketched in Fig. \ref{fig:model}. If the 
ionized material is flowing away from the ionization front at a speed of 
$\sim10\,$km$\,$s$^{-1}$ the thin molecular layer will be photo--evaporated in a 
time scale of
$\sim10^5$ years.

\section{Conclusions}

We have used the HHT Telescope to carry out observations of the 
$N=3\rightarrow2$ and new observations of the 
$N=2\rightarrow1$ lines of CN toward selected positions in the Trapezium region 
and in the molecular Ridge of the Orion A molecular cloud.
The main results are summarized as follow.

\begin{enumerate}

\item The hyperfine intensity ratio of just one transition of CN 
      does not give reliable estimate of the optical
      depth of the lines due to different excitation temperatures for 
different HF
      components.
      The ratios between HF components of the $N=3\rightarrow2$ line can be 
close to
      optically thin values even in the case of optically thick emission
      Observations of several HF groups with very different intensities or/and 
      observations
      of isotopic substitute of 
      CN are required to derive reliable opacities.

\item Our measurements of $N=3\rightarrow2$ line of CN confirm the presence of 
a high
      density (a few $10^6\,$cm$^{-3}$) molecular gas toward the Trapezium 
      region with also relatively large CN abundances ($\sim10^{-9}$). The 
H$_2$
      density in the Trapezium is a factor of 10 larger than in the Ridge.

\item The high density molecular gas towards the Trapezium is located in a 
      thin layer ($\sim 5\times10^{15}\,$cm). This thin molecular layer is 
      related with the material that 
      confines the far side of the H$\,$II region in the Trapezium region, 
      and it is ``static'' with respect to the molecular cloud. 

\item The kinematics of the ionized gas and the small thickness of the 
Trapezium layer
      can be understood in the framework of the champagne
      flows. The ``static'' molecular layer is in equilibrium with the H$\,$II 
region suggesting
      that evolved H$\,$II regions are not expanding in all directions as 
occurs for UC
      H$\,$II regions.

\end{enumerate}

\clearpage
\figcaption[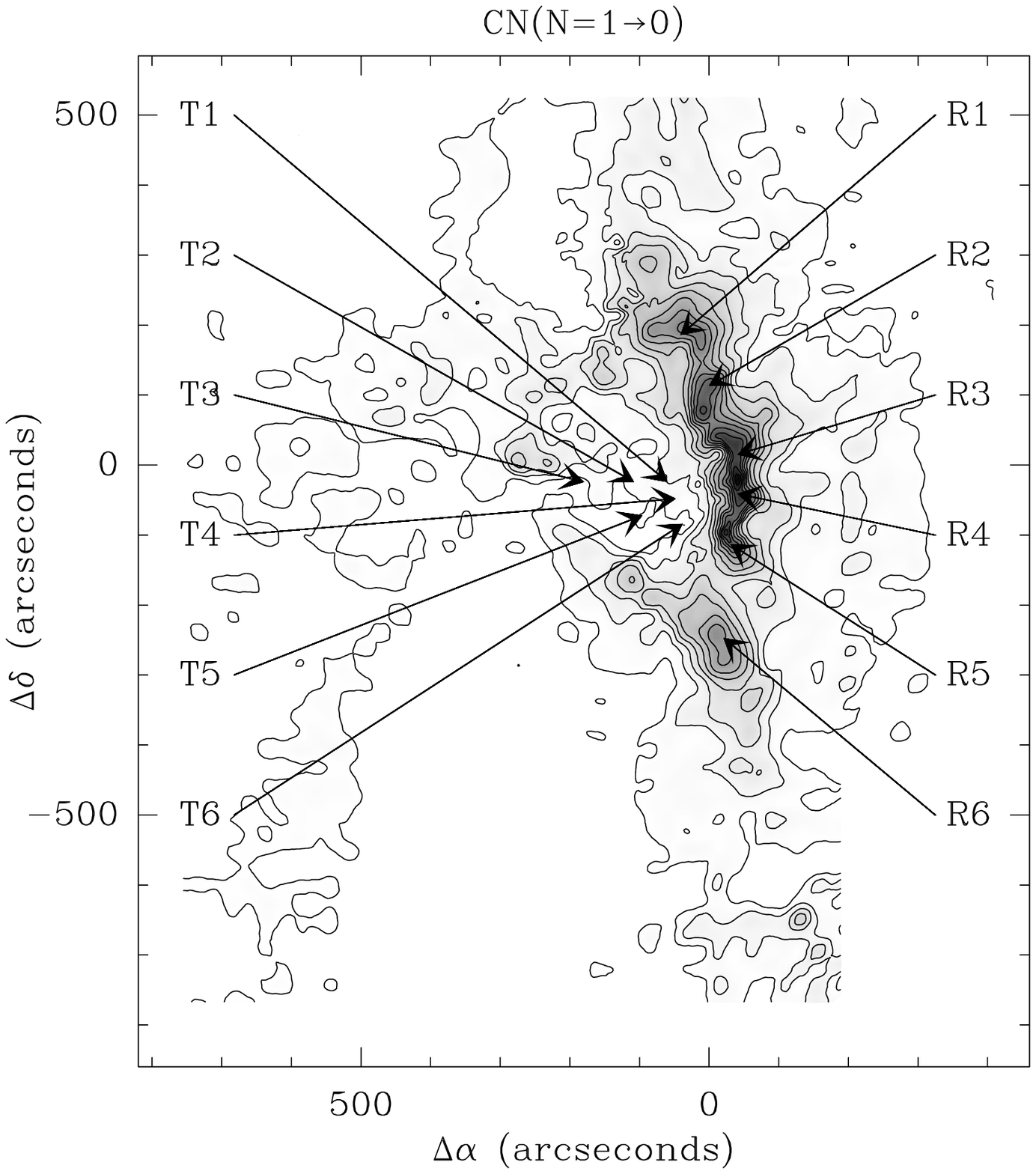]{Location of the positions observed in the CN $N=3\rightarrow2$ 
line in the
   Trapezium region (T1\ldots T6) and in the molecular Ridge (R1\ldots R6) 
superposed on
   the spatial distribution 
               of the CN $N=1\rightarrow0$ line 
               integrated intensity between 0 and 14$\,$km$\,$s$^{-1}$ 
               towards 
               Orion A from Rodr\'{\i}guez--Franco et al. (1998). The 
               offsets 
               are relative to the position of IRc2 
               ($\alpha$(1950)=5$^{{\rm h}}$ 32$^{{\rm m}}$ 47.0$^{{\rm s}}$, 
               $\delta$(1950)=--5$^{{\rm o}}$ 24$'$ 20.6$''$). First contour 
               level is 2$\,$K$\,$km$\,$s$^{-1}$ and the 
               interval between contours is 3.5$\,$K$\,$km$\,$s$^{-1}$. 
         \label{fig:mapa}}

\figcaption[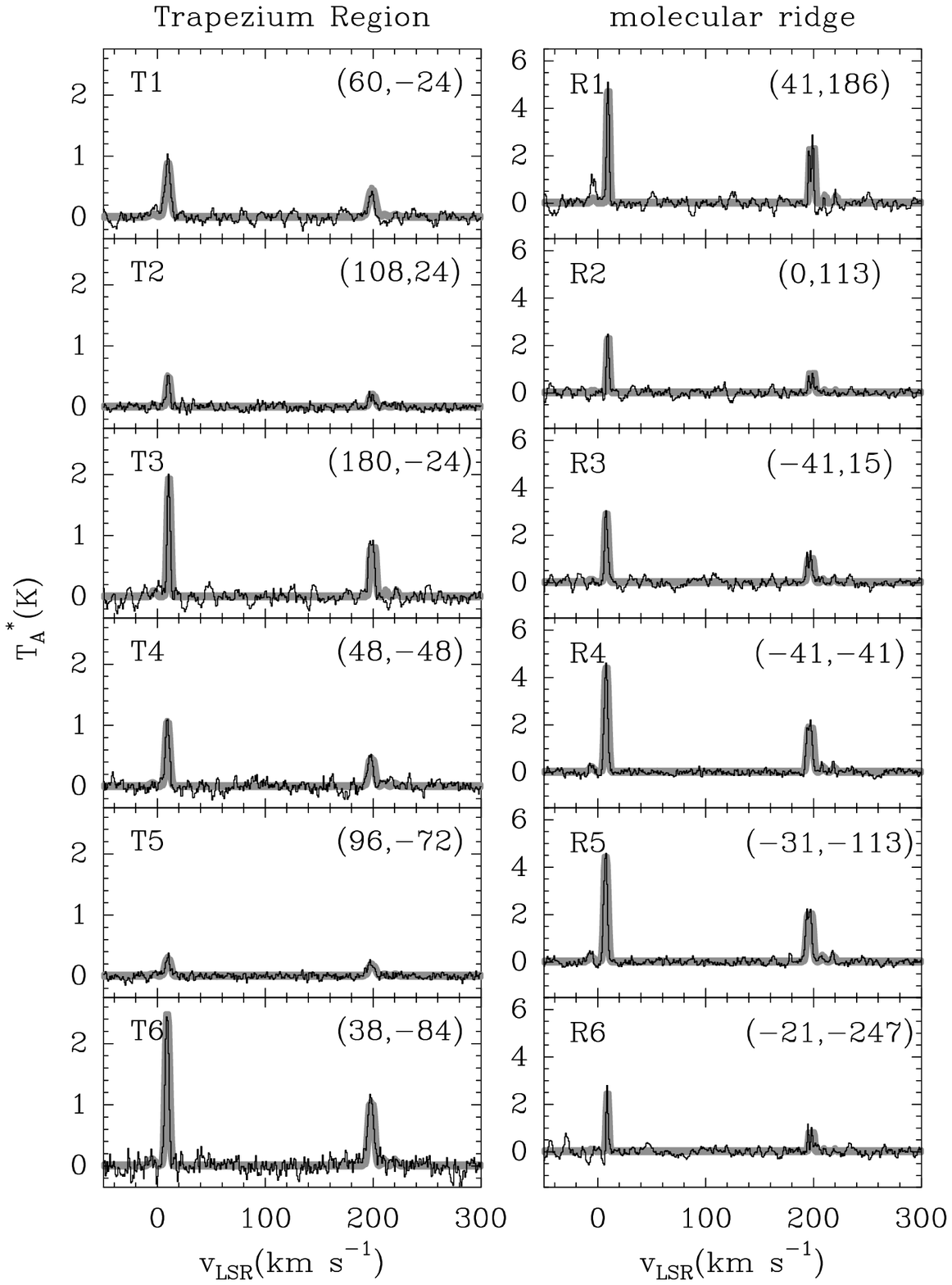]{Spectra (solid line) and fits (grey line) of the CN 
   $N=3\rightarrow2$ line taken towards the 
Trapezium
region
   (left panels) and towards the molecular Ridge (right panels). The position 
(offsets in
   arc-seconds respect the position of IRc2) where the
   spectra were taken is noted in the upper right corner
   of each box. In the upper left corner appear the nomenclature 
   used in Fig. 1.
         \label{fig:espec}}

\figcaption[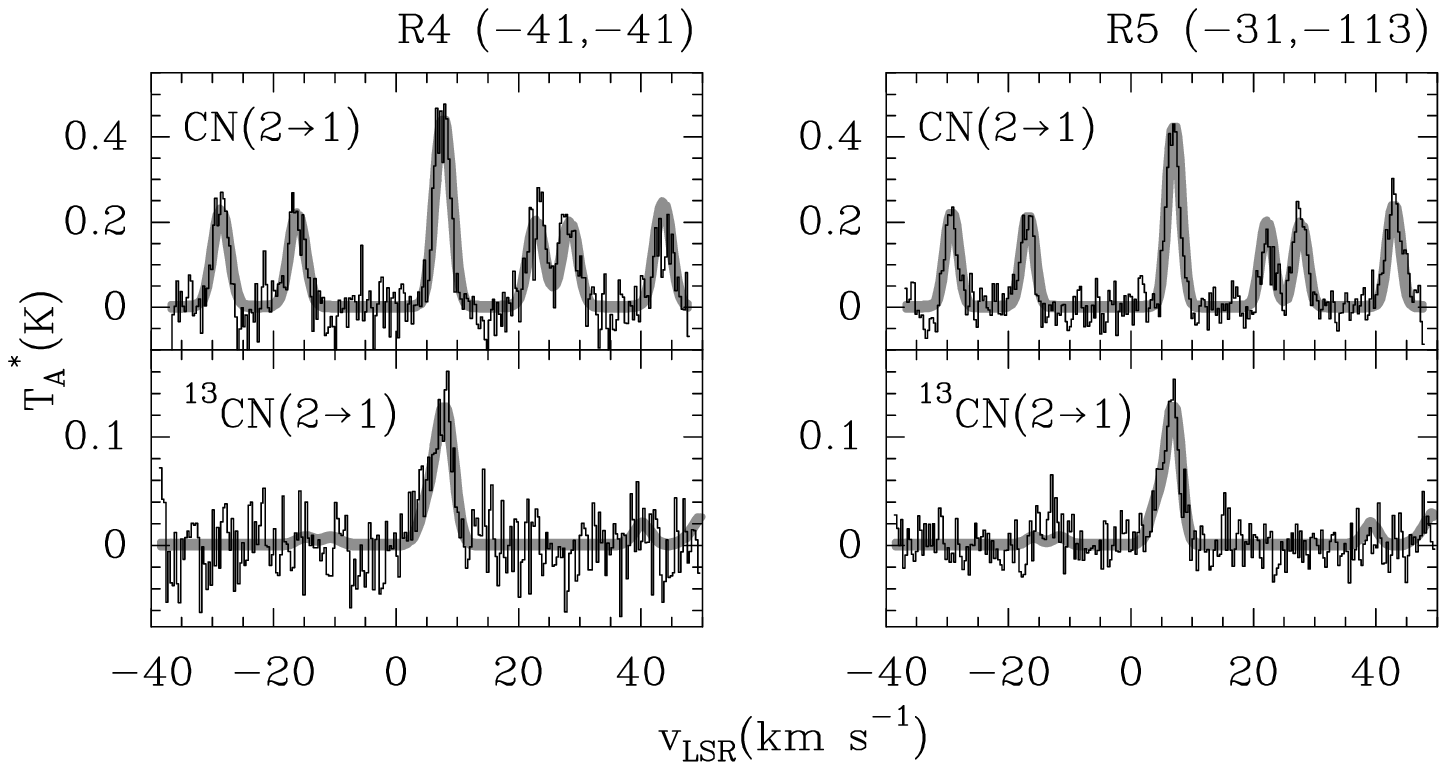]{Spectra (solid line) and fits (grey line)
   of the CN $N=2\rightarrow1$ and $^{13}$CN $N=2\rightarrow1$ line taken 
   towards two positions in the Ridge.
         \label{fig:iso}} 
	 
\figcaption[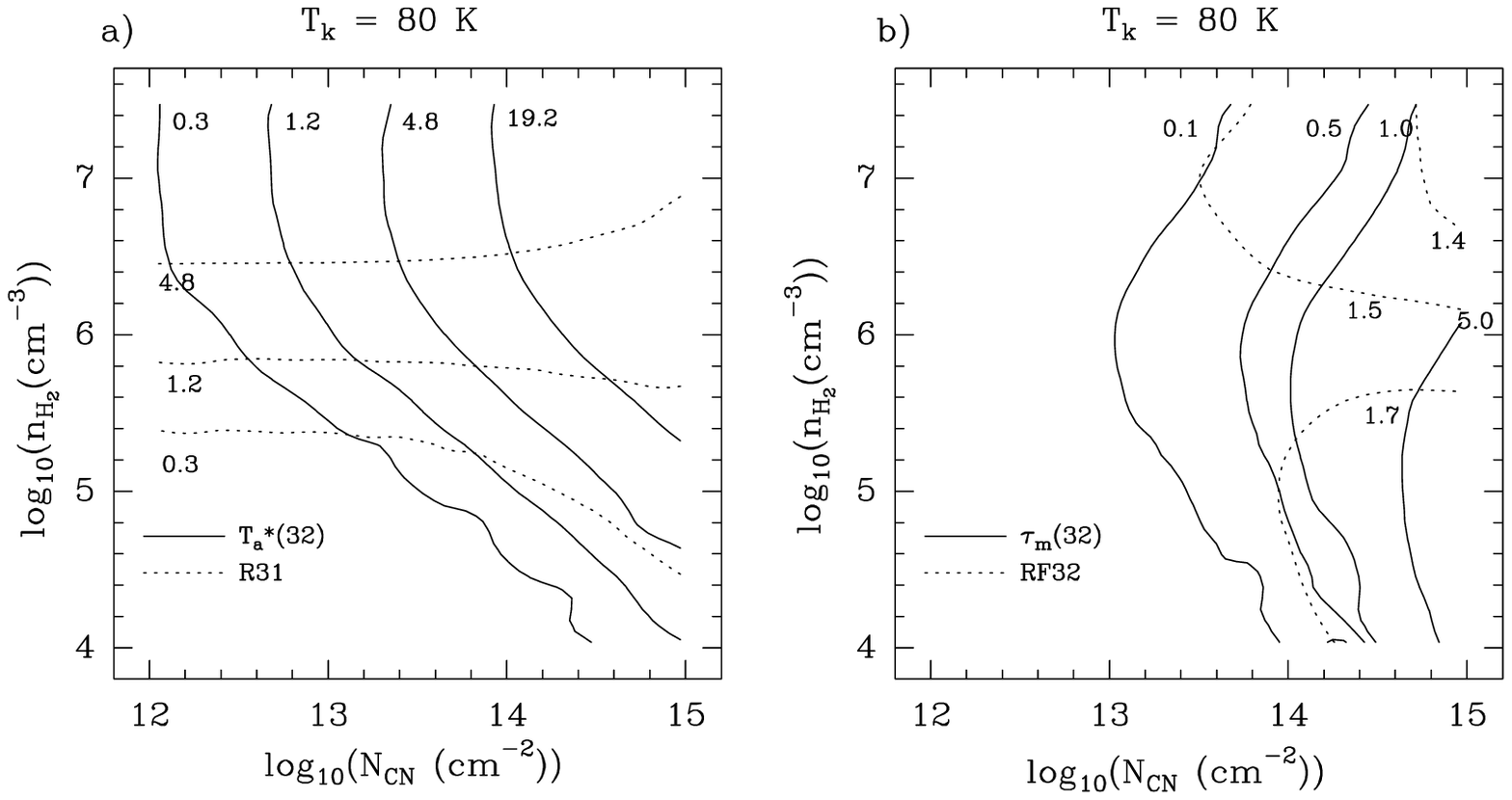]{LVG diagrams for a kinetic temperature of $80\,$K.
    {\bf a)} Solid lines represent the intensity in Kelvin of the main fine group 
    component of the 
    $N=3\rightarrow2$
    multiplet. Dotted lines, labeled R31,  represent the ratio between the intensity 
    of the
    main fine group of the $N=3\rightarrow2$ multiplet and the intensity of the main fine
    group of the $N=1\rightarrow0$ multiplet.
    {\bf b)} Solid lines represent the opacity of the main HF component of the 
     fine group component of the 
    $N=3\rightarrow2$
    multiplet. Dotted lines represent the ratio between the intensity of 
    the $J=7/2\rightarrow5/2$ and $J=5/2\rightarrow3/2$
    fine groups of the $N=3\rightarrow2$ multiplet
         \label{fig:lvg}}

\figcaption[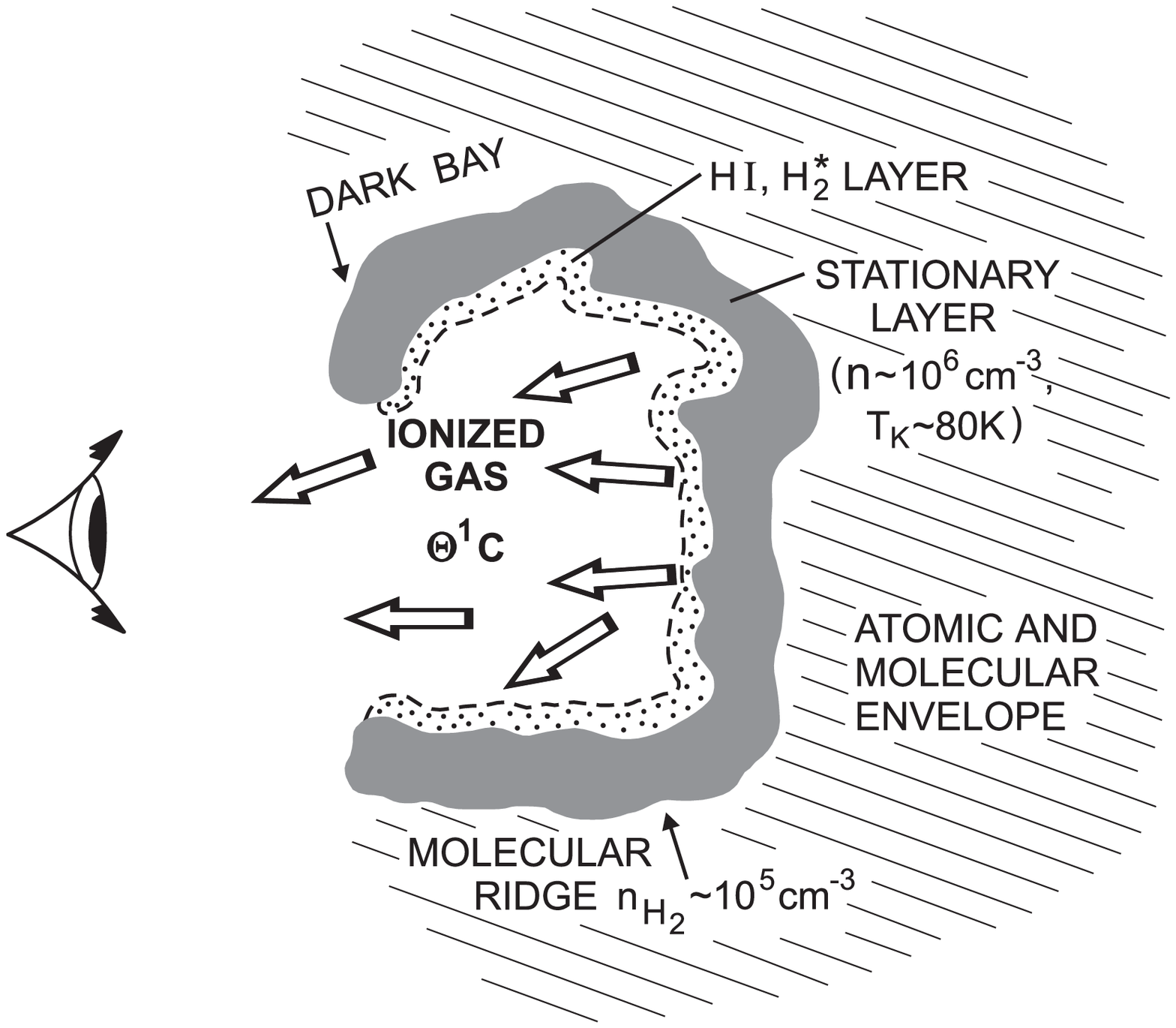]{Sketch of the of the distribution and kinematics of the ionized and
molecular gas towards the Trapezium. The sizes of the different layers are not
represented with the same scale and they have chosen to show the different 
components. \label{fig:model}}	 	 

\clearpage

\begin{deluxetable}{lccccccc}
\small
\tablecaption{Observational parameters of the CN emission. \label{tab:obs}}
\tablewidth{0pt}
\tablehead{
\colhead{Position} & \colhead{$T_a^*$(32)\tablenotemark{a}}   & \colhead{v$_{\rm LSR}$\tablenotemark{a}}   & 
\colhead{$\Delta v$\tablenotemark{a}} & 
\colhead{$T_a^*$(21)\tablenotemark{b}}  & \colhead{$T_a^*$($^{13}$21)\tablenotemark{b}} & 
\colhead{$T_a^*$(10)\tablenotemark{c}} & \colhead{RF(32)}\\
\colhead{($''$,$''$)} & \colhead{(K)}   & \colhead{(km$\,$s$^{-1})$}   & 
\colhead{(km$\,$s$^{-1}$)} & 
\colhead{(K)}  & \colhead{(K)} & \colhead{(K)} & \colhead{}
} 
\startdata
T1 (60,$-24$) & 0.9 & 9.5 & 5.9 & & & 0.4 & 1.6 \\
T2 (108,24) & 0.6 & 9.9 & 3.8 & & & $<0.4$ & 1.4 \\
T3 (180,$-24$) & 2.1 & 10.4 & 3.2& & & 1.2 & 1.3 \\
T4 (48,$-48$) & 1.1 & 9.2 & 4.1 & & & 1.3 & 1.6 \\
T5 (96,$-72$) & 0.3 & 9.7 & 5.3 & & & $<0.6$ & 0.9 \\
T6 (38,$-84$) & 2.6 & 8.8 & 3.9 & & & 0.6 & 1.3 \\
R1 (41,186) & 5.4 & 9.6 & 2.5 & & & 10.9 & 1.3 \\
R2 (0,113)  & 2.8 & 9.2 & 2.4 & & & 12.0 & 1.6 \\
R3 ($-41$,15) & 3.1 & 7.8 & 3.2 & & & 4.9 & 1.5 \\
R4 ($-41$,$-41$) & 4.7 & 7.6 & 3.6 & 0.46 & 0.12 & 7.5 & 1.3 \\ 
R5 ($-31$,$-113$) & 4.7 & 7.2 & 3.7 & 0.43 & 0.13 & 9.0 & 1.5 \\
R6 ($-21$,$-247$) & 3.1 & 8.6 & 1.7 & & & 6.2 & 1.4 \\
\enddata
\tablenotetext{a}{Fits to the  $F=9/2\rightarrow7/2$, $F=7/2\rightarrow5/2$, 
                  and $F=7/2\rightarrow3/2$ hyperfine components of the 
		  $N=3\rightarrow2$, $J=7/2\rightarrow5/2$ multiplet}
\tablenotetext{b}{Fits to the  $F=3/2\rightarrow3/2$
                  hyperfine component of the $N=2\rightarrow1$,
		  $J=3/2\rightarrow3/2$ multiplet}
\tablenotetext{c}{Fits to the $F=5/2\rightarrow3/2$ hyperfine component of the 
                  $N=1\rightarrow0$, $J=3/2\rightarrow1/2$ multiplet (from 
		  Rodr\'{\i}guez$-$Franco et al. 1998)}

\end{deluxetable}

\clearpage
\begin{deluxetable}{lcccccc}
\small
\tablecaption{Derived parameters of the CN emission. \label{tab:derived}}
\tablewidth{0pt}
\tablehead{
\colhead{Position} & \colhead{$N({\rm CN}$)}   & \colhead{$N({\rm H}_2)$}   & 
\colhead{$n_{{\rm H}_2}$} & 
\colhead{$\chi$(CN)}  & \colhead{depth\tablenotemark{e}} & 
\colhead{$N(^{13}{\rm CN})$} \\
\colhead{($''$,$''$)} & \colhead{($\times10^{13}\,$cm$^{-2}$)}   & \colhead{($\times10^{22}\,$cm$^{-2}$)}   & 
\colhead{($\times10^{5}\,$cm$^{-3}$)} & 
\colhead{($\times10^{-9}$)}  & \colhead{($\times10^{16}\,$cm)} & \colhead{($\times10^{13}\,$cm$^{-2}$)}
} 
\startdata
T1 (60,$-24$) & 8.0 & 2.59\tablenotemark{a} & 6 & 3.1 & 4.31 &  \\
T2 (108,24) & 2.2 & 0.86\tablenotemark{a} & 12.5 & 2.5 & 0.69  & \\
T3 (180,$-24$) & 3.0 & 0.86\tablenotemark{a} & 50 & 3.5 & 0.17 & \\ 
T4 (48,$-48$) & 3.3 & 3.46\tablenotemark{a} & 6 & 0.9 & 5.77 & \\
T5 (96,$-72$) & 1.7 & 0.43\tablenotemark{a} & 10 & 3.9 & 0.43  & \\
T6 (38,-84) & 4.3 & 0.86\tablenotemark{a} &  70 & 4.9 & 0.12 & \\ 
R1 (41,186) & 245 &  & 1 & &  &  \\
R2 (0,113) & 240 & 6.06\tablenotemark{b} & 0.31 & 39 & 195  & \\ 
R3 ($-41$,15) & 30  & 10.4\tablenotemark{c} & 2 & 2.8 & 52.0 & \\ 
R4 ($-41$,$-41$) & 97 & 12.1\tablenotemark{c} & 1 & 8.0 & 121 & 1.3 \\
R5 ($-31$,$-113)$ & 75 & 12.1\tablenotemark{c} & 2 & 6.2 & 60.7 & 1.0 \\
R6 ($-21$,$-247$) & 90 & 11.2\tablenotemark{b} & 0.7 & 8.0 & 160 &   \\

\enddata
\tablenotetext{a}{From $^{13}$CO data (White \& Sandell, 1995)}
\tablenotetext{b}{From $^{13}$CO and C$^{18}$O data (White \& Sandell, 1995)}
\tablenotetext{c}{From C$^{18}$O data (White \& Sandell, 1995)}
\tablenotetext{d}{Assumed $n_{{\rm H}_2}$}
\tablenotetext{e}{From the $N_{{\rm H}_2}$ in column 3 and the $n_{{\rm H}_2}$ in column 4}

\end{deluxetable}

\clearpage 
\begin{figure} 
\plotone{fcn1.ps}
\end{figure}
\label{fig:mapa}

\clearpage

\begin{figure}
\plotone{fcn2.ps}
\end{figure}

\clearpage

\begin{figure}
   \plotone{fcn3.ps}
\end{figure}
 
\clearpage

\begin{figure}
\plotone{fcn4.ps} 
\end{figure}

\clearpage

 \begin{figure}
\plotone{fcn5.ps}
  \end{figure}


\begin{thebibliography}{}
   \bibitem[De Pree 1995]{DePree95} De Pree, C.G., Rodr\'{\i}guez, L.F., Goss, W.M., 1995,
              Rev. Mex. A.A 31, 39.
              
   \bibitem[Churchwell 1979]{Chur79} Churchwell, E., 1990, A\&AR 2, 79.

   \bibitem[Dreher 1982]{Dreher82} Dreher, J.W., Welch, W.J., 1981, ApJ 245, 857.
   
   \bibitem[Dreher 1984]{Dreher84} Dreher, J.W., Johnston, K.J., Welch, W.J., Walker,
              R.C., 1984, ApJ 283, 632.

   \bibitem[Dyson1995]{Dyson95} Dyson, J.E., Williams, R.J.R., Redman, M.P., 1995, 
              MNRAS 277, 700.
              
   \bibitem[Fuente 1993]{Fuente93} Fuente, A., Mart\'{\i}n--Pintado, J., Bachiller, R., 
              Cernicharo, J., 1993, A\&A 276, 473.
   
   \bibitem[Fuente 1995]{Fuente95} Fuente, A., Mart\'{\i}n--Pintado, J., Gaume, R., 1995,  
              ApJLetters 442, L33.     
              
   \bibitem[Gaume 1996]{Gaume96} Gaume, R.A.. Wilson, T.L., Johnston, K.J., 1996 ApJ 457, L47.
   
   \bibitem[Groesbeck 1994]{Groesbeck94} Groesbeck, T.D., Phillips, T.G., Blake, G.A., 1994,
              ApJS 94, 147.
   
   \bibitem[Hollenbach 1994]{Hollenbach94} Hollenbach, D., Johnstone, D., Lizano, S., Shu, F., 1994,
              ApJ 428, 654.
	      
   \bibitem[Hollenbach 1997]{Hollenbach97} Hollenbach, Tielens, A. 1997 ARA\&A 35, 179.  
               
   \bibitem[Howe 1993]{Howe93} Howe, J.E., Jaffe, D.T., Grossman, E.N., Wall, W.F., Mangum, 
              J.G., Stacey, G.J., 1993, ApJ 410, 179.
              
   \bibitem[Jansen 1995]{Jansen95} Jansen, D.J., Spaans, M., Hogerheijde, M.R., 
              van Dishoeck, E.F., 1995, A\&A 303 541. 

   \bibitem[Lizano 1996]{Lizano96} Lizano, S., Cant\'o, J., Garay, G., Hollenbach, D.J., 
              1996, ApJ 468, 739.
   
   \bibitem[Luhman 1994]{Luhman94} Luhman, M.L., Jaffe, D.T., Keller, L.D., Soojong Pak, 
              1994, ApJLet 436, L185.     

   \bibitem[Jaffe 1999]{Jaffe99} Jaffe, D.T., Mart\'{\i}n--Pintado, J., 1999, ApJ 520, 162.
   
    \bibitem[O'Dell 1993]{O'Dell93} O'Dell, C.R., Valk, J.H., Wen, Z., Meyer, D.M., 1993, 
              ApJ 403, 678.  

   \bibitem[Pauls 1983]{Pauls83} Pauls, A., Wilson, T.L., Bieging, J.H., Martin, R.N., 1983 
              A\&A  124, 23.

   \bibitem[Reid 1980]{Reid80} Reid, M.J., Haschick, A.D., Burke, B.F., Moran, J.M.,
              Johnston, K.J., Swenson, G.W. Jr., 1980, ApJ 239, 89.       
             
   \bibitem[Rodr\'{\i}guez--Franco 1992]{Rodri92} Rodr\'{\i}guez--Franco, A., Mart\'{\i}n--Pintado, J.,
              G\'omez--Gonz\'alez, J., Planesas, P., 1992 A\&A 264, 592.   
                               
   \bibitem[Rodr\'{\i}guez--Franco 1998]{Rodri98} Rodr\'{\i}guez--Franco, A., Mart\'{\i}n--Pintado J., Fuente, A, 
              1998, A\&A 329, 197

   \bibitem[Schilke 1997]{Schilke97} Schilke, P., Groesbeck, T.D., Blake, G.A., Phillips, T. G.,    
              1997 ApJS 108, 301.
  
   \bibitem[Sternberg 1995]{Sternberg95} Sternberg, A., Dalgarno, A., 1995, ApJss 99, 565.  
              
   \bibitem[Tenorio--Tagle 1979]{Tenorio79} Tenorio--Tagle, G., 1979, A\&A 71, 59.
   
   \bibitem[Van Buren 1990]{VanBuren90} Van Buren, D., Mac Low, M. M., Wood, D.O.S.,
              Churchwell, E., 1990, ApJ 353, 570.
	      
   \bibitem[van der Werf 1989]{vanderWerf89} van der Werf, P.P., Goss, W.M., 1989, A\&A 224, 209.
   
   \bibitem[White 1995]{White95} White G.J., Sandell G., 1995, A\&A 299, 179.   

   \bibitem[Wilson 1986]{Wilson86} Wilson, T.L., Serabyn, E., Henkel, C., Walmsley, C.M., 1986, 
              A\&A 158, L1.
   
   \bibitem[Wilson 1997]{Wilson97} Wilson, T.L., Filges, L., Codella, C., Reich, W., Reich, P.,  
              1997, A\&A 327, 1177.
              
   \bibitem[Wilson 1999]{Wilson99} Wilson, T.L., 1999, Rep. Prog. Phys 62, 143.
   
   \bibitem[Wilson 2000]{Wilson00} Wilson, T.L., Muders, D., Kramer, C., Henkel, C., 2000,
              ApJ, submited.

   \bibitem[Wood 1989]{Wood89} Wood, D.O.S., Churchwell, E., 1989, ApJSupp 69, 831.

   \bibitem[Zuckerman 1973]{Zuckerman73} Zuckerman, B., 1973, ApJ 183, 863.

\end{thebibliography}
\end{document}